\documentclass[prd,
               twocolumn,
               showpacs,nofootinbib,amsmath,amssymb]{revtex4}
\usepackage{graphicx}

\newcommand{\cD}{\mathcal{D}}
\newcommand{\half}{\frac{1}{2}}
\newcommand{\bra}[1]{\langle #1 |}
\newcommand{\ket}[1]{| #1 \rangle}

\newcommand{\e}{\mathrm{e}}
\newcommand{\const}{\mathrm{const}}

\newcommand{\dm}{\partial_\mu}

\newcommand{\Dm}{D_\mu}
\newcommand{\atanh}{\mathop\mathrm{atanh}}
\newcommand{\Pleft}{\frac{1+\gamma^5}{2}}
\newcommand{\Pright}{\frac{1-\gamma^5}{2}}
\newcommand{\Ker}{\mathop\mathrm{Ker}}
\newcommand{\identity}{\openone}

\begin{document}

\title{Can an odd number of fermions be created due to chiral anomaly?}
\author{F. Bezrukov}\email{Fedor.Bezrukov@epfl.ch}
\author{Y. Burnier}\email{Yannis.Burnier@epfl.ch}
\author{M. Shaposhnikov}\email{Mikhail.Shaposhnikov@epfl.ch}
\affiliation{
  Institut de Th\'eorie des Ph\'enom\`enes Physiques,\\
  \'Ecole Polytechnique F\'ed\'erale de Lausanne,\\
  CH-1015 Lausanne, Switzerland    
}
\date{December 13, 2005}

\begin{abstract}
  We describe a possibility of creation of an odd number of
  fractionally charged fermions in 1+1 dimensional Abelian Higgs
  model.  We point out that for 1+1 dimensions this process does not
  violate any symmetries of the theory, nor makes it mathematically
  inconsistent. We construct the proper definition of the fermionic
  determinant in this model and underline its non-trivial features
  that are of importance for realistic 3+1 dimensional models with
  fermion number violation.
\end{abstract}
\pacs{11.15.Kc, 11.30.Fs, 11.30.Rd}

\maketitle

\section{Introduction}

It is well known that many gauge theories with nontrivial topological
structure allow for violation of fermion number $N_F$.  A familiar example
is just the Standard Model. The instanton processes in it lead to
non-conservation of $N_F$ by an even number, equal to four
times the number (three) of fermionic generations. A model with
SU(2) gauge group and just one fermion in fundamental representation
would predict, na\"ively, the  processes that change the vacuum
topological number by one which would lead to creation of just one
fermion. This type of process contradicts to quite a number of
principles of quantum field theory, such as spin-statistics relation,
Lorentz invariance, etc. A resolution of the paradox is known:  this
model turns out to be mathematically inconsistent, because of so
called global Witten anomaly~\cite{Witten:1982fp}. The Witten anomaly
is connected with the topological fact that the fourth (four comes
from the number of space-time dimensions) homotopy group
$\pi^4(SU(2))=\mathbb{Z}_2$ is non-trivial.  This makes it impossible
to define a measure in the functional integral over fermion fields in
the models with an odd number of fermionic doublets. The anomaly
disappears if the number of fermionic doublets is even, but then
fermions are always created in pairs.

Clearly, the Witten consistency condition does depend on
the dimensionality of space-time and may change if the number of
dimensions is not equal to four. For example, in two dimensional
Abelian gauge theories, the topological considerations are different.
The corresponding homotopy group $\pi^2(U(1))=0$ is trivial and the
fermionic measure can be defined properly\footnote{Strictly speaking
  Witten like anomaly can occur even in theories with trivial
  $\pi_{d+1}$ homotopy group that allow one fermion creation,
  see \cite{Witten:1985xe}.  The argument
  there is inaplicable to 1+1--dimensional case see discussion in the
  section \ref{wittenanomaly}.}. So one may expect
existence of processes with one fermion creation in 1+1 dimensions.

This article is devoted to the demonstration that this effect really
takes place in 1+1 dimensional models, specifically in an Abelian
Higgs model with a chirally charged fermion of half integer charge. 
It will be shown that the creation of one fermion in 1+1 dimensions does not
contradict neither to Lorentz symmetry, nor the calculation of the
cross section of such a process leads to some unexpected
cancellations. 

There are generally two methods with which one can see that the
processes with creation or decay of one fermion can take place.  We
will use both of them in this work. The first one is the analysis of
fermion level crossing in the topologically nontrivial background
\cite{Callan:1977gz,Kiskis:1978tb,Nielsen:1983rb,Ambjorn:1983hp}.
This picture is straightforward and very intuitive, but it does not
allow (at least easily) for calculation of the probability or cross-
section of the corresponding process.

The second method uses perturbation theory in instanton background.
It was widely used in the calculation of baryon number violating
processes
\cite{Krasnikov:1979bh,Krasnikov:1978dg,Ringwald:1990ee,Espinosa:1990qn}.
The exponent of the probability is easily obtained in this approach,
but the preexponential factor is much harder to calculate.  For the
theories with chiral fermions it was estimated before only using
dimensional considerations for part of the computation.  The correct
definition of the preexponential factor (or, equivalently, the
fermionic determinant) is nontrivial.  This was noted for example in
\cite{Nielsen:1976hs,Stora:1976kd}.  In this article we construct a
consistent way to calculate the preexponent in theories with chiral
fermions.  It is important to note that the same problem also occurs
in the usual 4-dimensional electroweak theory, where a similar
procedure should be used to obtain the correct prefactor in the
instanton transition probability.

The paper is organized as follows.  In Section~\ref{sec:LIandSS} we
analyze the general properties of two-dimensional models, namely
Lorentz transformation properties of the Greens' functions and absence
of superselection rules and Witten like global anomalies.  These
properties differ from higher dimensional ones and lead to possibility
of one fermion creation.  Section~\ref{sec:levelcrossing} describes
the model we study and its vacuum structure. We explain here the
creation of one fermion using level crossing approach.  Instanton
calculation of the cross section is given in the
Section~\ref{sec:instantons}.  Conclusions are presented in the
Section~\ref{sec:concl}.  In Appendices
\ref{sec:ave}--\ref{sec:detcalculation} we describe some technical
details of computations.

\section{Lorentz Invariance and Superselection Rules}
\label{sec:LIandSS}

\subsection{Lorentz invariant one fermion Greens' functions}

Usually, processes with an odd number of fermions participating in the reaction are automatically
forbidden by Lorentz symmetry.  Let us show that in 1+1 dimensions it
is not the case, i.e.\ Lorentz invariant Greens' functions with one
fermion can be non-trivial.

Two dimensional spinors transform under a Lorentz boost $\Lambda$ with
rapidity $\beta$ in the following way,
\begin{multline}\label{SpinorLorentz}
  \Psi(x) \to \Psi'(x) = \Lambda_\half\Psi(\Lambda^{-1}x) \\
  = \e^{-\frac{\beta}{2}\gamma^5}\Psi(\Lambda^{-1}x)
  = \begin{pmatrix}
      \e^{-\frac{\beta}{2}} \Psi_L(\Lambda^{-1}x) \\
      \e^{\frac{\beta}{2}} \Psi_R(\Lambda^{-1}x)
    \end{pmatrix}
  \;.
\end{multline}
Requirement of the Lorentz invariance of the simple Green's function
with one fermion has the following form, supposing that the vacuum is
Lorentz invariant
\begin{align*}
  G(x;y) = \bra{0} \Psi(x) \phi(y) \ket{0}
  &= \bra{0} U^{-1}(\Lambda) \Psi(x) \phi(y) U(\Lambda) \ket{0}
  \\
  &= \bra{0} \Lambda_\half\Psi(\Lambda^{-1}x) \phi(\Lambda^{-1}y)
     \ket{0}
  \;.
\end{align*}
Moving $y$ to the coordinate origin, $y=0$, we get for the left and
right components the equations (writing space and time dependence
explicitly)
\begin{align*}
  G_L(x^0,x^1;0,0) &=
    \begin{array}[t]{r@{}l}
      \e^{-\frac{\beta}{2}}
      G_L(&x^0\cosh\beta-x^1\sinh\beta,\\
          &x^1\cosh\beta-x^0\sinh\beta;0,0) \;,
    \end{array}
  \\
  G_R(x^0,x^1;0,0) &=
    \begin{array}[t]{r@{}l}
      \e^{\frac{\beta}{2}}
      G_R(&x^0\cosh\beta-x^1\sinh\beta,\\
          &x^1\cosh\beta-x^0\sinh\beta;0,0) \;.
    \end{array}
\end{align*}
These equations allow solution
\begin{align*}
  G_{L,R}(x^0,x^1;0,0) & =
  \exp\left[\pm\half\atanh\left(-\frac{x^0}{x^1}\right)\right]
    f_{L,R}(x_\mu x^\mu)
  \\
  & = \sqrt[4]{\frac{x^0\mp x^1}{x^0\pm x^1}}
    \,f_{L,R}(x_\mu x^\mu)
\end{align*}
with arbitrary functions $f_{L,R}$.

Similar solutions can be found also for more complicated Greens'
functions.  So, in 1+1 dimensions, thanks to the simple form of
Lorentz transformation~(\ref{SpinorLorentz}), Greens' functions
containing an odd number of fermion fields are not necessarily equal
to zero.

\subsection{Absence of superselection rules}
\label{sec:appLorentz}

We follow here the arguments given in \cite{Andreev:2003nf}.  In 3+1
dimensions a coherent superposition of states with even $|even\rangle$
and odd $|odd \rangle$ numbers of fermions is incompatible with
Lorentz invariance.  More precisely, a state with an odd number of
fermions is multiplied by $(-1)$ under rotation of $2\pi$ of the
coordinate system around any axis and under double application of time
reversal. Then clearly superpositions of even and odd states would
change under previously mentioned transformations which coincide with
identity:
\begin{equation*}
  |even\rangle+|odd \rangle \overset{2\pi~\mathrm{
  rotation}}\longrightarrow |even\rangle-|odd \rangle
  \;.
\end{equation*}
In $1+1$ dimensions the Lorentz group consists of a boost only.  There
is no rotation, and double application of time reversal does not give
a factor $(-1)$.  Indeed, time reversal in two dimensions is:
\begin{equation*}
  T=T_0K \mathcal{T} =i\gamma^1K\mathcal{T}
  \;,
\end{equation*}
where the operator $\mathcal{T}$ changes $t \to -t$, $K$ performs the
complex conjugate and $T_0=i\gamma^1$ is a matrix in spinor space
chosen so that the Dirac equation remains unchanged under time
reversal. Note that $i\gamma^1$ is real and symmetric. Then
\[
  T^2=i\gamma^1K i\gamma^1 K = (i\gamma^1)^2=\identity
  \;.
\]
Parity transformation can also be defined not to give factor $(-1)$
after double application.

So there are no superselection rules contradicting with considering
configurations with odd number of fermions in 1+1 dimensions.

\subsection{Absence of Witten anomaly}
\label{wittenanomaly}

As we allready mentioned in the introduction, there is a global Witten
anomaly in $d$--dimensional gauge theories with gauge group $G$ and
nontrivial $\pi_d(G)$.  This is not the case for our model, because
$\pi_2(U(1))$ is zero.  But there is a rather simple argument by
Goldstone, present in \cite{Witten:1985xe}, that relates the existence
of the global anomaly to the possibility of creation of odd number of
fermions in the instanton processes (or to odd number of fermion zero
modes in the instanton background).  The argument is rather short and
nice and we will present it here.

Let us suppose we have a gauge theory with an Yang--Mills instanton.
Let us call $\pi$ the gauge transformation associated with the
instanton (which transforms between the vacua that are connected by
the instanton), and $\Lambda$ the corresponding operator acting on the
quantum Hilbert space.  The Gauss law requires that all gauge or
coordinate transformations that can be connected continuously with
identity leave the physical states invariant.  $\Lambda$ is not
constrained by Gauss law, since $\pi$ is a topologically nontrivial
transformation, and is generally equal to $\e^{-i\theta}$, where
$\theta$ is some phase.

Now, if the instanton is associated with odd number of zero modes, we
have
$(-1)^F\Lambda(-1)^F=-\lambda$, where $(-1)^F$ counts the fermion
number $\!\!\mod2$.

Let us now take an generator $J$ of spatial rotations along some axis,
and construct the operator
\[
  G_s = \pi^{-1}\exp(-isJ)\pi\exp(isJ) \;.
\]
By construction $G_0=1$, therefore Gauss law predicts that all
physical states $G_s\ket{\mathrm{physical}}$ should be identical.  However,
$G_{2\pi}=\pi^{-1}(-a)^F\pi(-1)^F=-1$.  This means that the Hilbert
space does not exist, which is a synonym of a global anomaly
\cite{Witten:1982fp}.

However, in our case this argument fails because of absence of spatial
rotations.  This means that the two dimensional theories should be
free of global anomalies, and this should be the only case free of
global anomalies allowing one fermion creation.

\section{The Model and Level Crossing Description}
\label{sec:levelcrossing}

\subsection{The Model}

We are analyzing a chiral Abelian Higgs model in 1+1 dimensions with
one fermion of a half charge.  The Lagrangian of the model is
\begin{align}\label{Lm}
  \mathcal{L} = &
    -\frac{1}{4}F^{\mu\nu}F_{\mu\nu}
    +\frac{1}{2}\left|\Dm\phi\right|^2
    +i\overline\Psi\gamma^\mu\Dm\Psi
  \notag \\
  & -\frac{\lambda}{4}(|\phi|^2-v^2)^2
  \notag \\
  & +if\left(\overline{\Psi} \Pleft \Psi\phi^*
      -\overline{\Psi} \Pright \Psi\phi\right)
  \;,
\end{align}
where covariant derivatives are
\begin{align*}
  \Dm\phi&=(\dm-ieA_\mu)\phi
  \;, &
  \Dm\Psi&=(\dm-i\frac{e}{2}\gamma^{5}A_{\mu})\Psi
  \;.
\end{align*}
We use the two dimensional Dirac matrices representation
\begin{equation*}
  \gamma^{0}=\begin{pmatrix}
    0 & -i \\ 
    i & 0
  \end{pmatrix}
  \,,\;
  \gamma^{1}=\begin{pmatrix}
    0 & i \\ 
    i & 0
  \end{pmatrix}
  \,,\;
  \gamma^5=\gamma^0\gamma^1=\begin{pmatrix}
    1 & 0 \\
    0 & -1
  \end{pmatrix}
  \,,
\end{equation*}
and Dirac conjugate spinor is $\overline\Psi=\Psi^\dagger\gamma^0$.

The charges of the left and right-handed components of the fermion
$\Psi=\left(\begin{smallmatrix}\Psi_L\\
\Psi_R\end{smallmatrix}\right)$ differ by a sign,
$e_L=-e_R=\frac{e}{2}$.  This model has been studied as a toy model
for fermionic number non-conservation in electroweak theory in a
number of papers, see,
e.g.~\cite{Grigoriev:1988bd,Bochkarev:1987wg,Grigoriev:1989je,
Grigoriev:1989ub,Bochkarev:1989vu,Baacke:1994bk,Kripfganz:1989vm}.

The particle spectrum consists of a Higgs field with mass
$m_H=\sqrt{2\lambda}v$, a vector boson of mass $m_W=ev$, and a Dirac
fermion acquiring a mass $F=fv$ via Higgs mechanism.  The model is
free of gauge anomaly.  There is, however, a chiral anomaly leading to
non-conservation of fermionic current,
\[ 
  J_\mu = J_\mu^L+J_\mu^R
  = \overline{\Psi}\gamma_\mu\Psi
  \;,
\]  
with a divergence given by 
\begin{equation}\label{curentanom}
  \dm J_\mu = \partial_\mu J_\mu^L+\partial_\mu J_\mu^R
  = -q
  \;,
\end{equation} 
where $q=\frac{e}{4\pi}\varepsilon_{\mu\nu}F_{\mu\nu}$ is the winding
number of the gauge fields configuration.  This immediately leads to
the conclusion that in topologically nontrivial backgrounds one can
get creation of only one fermion.

The simplest description of fermion number violating processes in
gauge theories is obtained from the analysis of the fermionic level
structure in nontrivial external bosonic fields.  First, we have to
describe the level structure in different topological vacua, and then
analyze the level crossing picture in gauge field background
interpolating between vacua with topological numbers different by one.

To clarify the topological structure we will insert the system in a
finite box of length $L$ with periodic boundary conditions. At the
end, the parameter $L$ can be taken to infinity to recover the
infinite space results.



\subsection{Gauge transformations and fermion spectrum}

Zero energy configurations of the gauge and Higgs fields are
obtained by gauge transformations from the trivial vacuum state
\begin{equation*}
  \phi^\mathrm{vac} = \e^{i\alpha(x)}v
  \;,\quad
  A_\mu^\mathrm{vac} = \frac{1}{e}\dm\alpha(x)
  \;.
\end{equation*}
These configurations will be called bosonic vacua. In infinite space,
or in finite space with periodic boundary conditions for the bosonic
fields, the configurations are divided into topological sectors,
labeled by the topological number
$n=\frac{1}{2\pi}(\alpha(\infty)-\alpha(-\infty))$.

\begin{figure} 
  \begin{center}
    \includegraphics[width=\columnwidth]{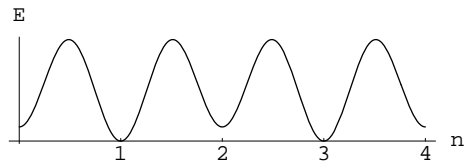} 
    \caption{Picture of the fermionic energy in different bosonic configurations. Bosonic vacua with odd $n$ have a slightly different energy}
    \label{vacua} 
  \end{center} 
\end{figure}

Let us see what happens with fermions when we apply (large) gauge
transformations changing the topological number of the vacuum.  To
leave the Lagrangian~(\ref{Lm}) invariant fermion fields should
transform as
\begin{equation*}
  \Psi \rightarrow \e^{i\alpha (x)\frac{\gamma _{5}}{2}}\Psi
  \;,\quad
  \overline{\Psi} \rightarrow
    \overline{\Psi} \e^{i\alpha (x)\frac{\gamma _{5}}{2}}
  \;.
\end{equation*}
The fractional fermion charge leads here to some complications.  For
gauge transformations with odd $n$ the transformation spoils the
boundary conditions for the fermion wave function $\Psi$.  So, at
least in finite size system, fermion spectra in bosonic vacua with
even and odd topological numbers are different. As a result, the
energies of the lowest states with odd and even topological numbers
are different as well. In other words, the bosonic vacuum states with
even $n$ have higher energy than the states with odd $n$ (see
Appendix~\ref{sec:ave}) and therefore are not the true vacua of the
theory\footnote{This difference disappears in the limit of infinite
space, see Appendixes~\ref{sec:ave} and \ref{sec:appFnumber}.} (see
Fig.~\ref{vacua}).

Let us analyze this feature in more detail.

The fermionic equation of motion is:
\begin{equation*}
  \left[i\partial_0-H_D\right]\Psi=0
  \;,
\end{equation*}
with Dirac Hamiltonian
\begin{equation*}
  H_D=\left(\begin{array}{cc} 
    -i\partial_1-\frac{e}{2}A_1 & f\phi \\
    f\phi^* & i\partial_1-\frac{e}{2}A_1 
  \end{array}\right)
  \;.
\end{equation*}
In trivial background ($A_\mu=0$, $\phi=v$) in a box of size $L$ with
periodic boundary conditions positive and negative energy fermionic
solutions have the form
\begin{align}\label{n=0}
  \Psi_{+} &=
    \e^{-iE_lt}\begin{pmatrix}
      \e^{i\frac{2\pi l}{L}x}F \\
      \e^{i\frac{2\pi l}{L}x}(E_l-k_l)
	       \end{pmatrix}
  \;,\\
  \Psi_{-} &= \hphantom{\scriptstyle -}
    \e^{iE_lt} \begin{pmatrix}
      \e^{i\frac{2\pi l}{L}x}(E_l-k_l) \\
      -\e^{i\frac{2\pi l}{L}x}F
	       \end{pmatrix}
  \;, \notag
\end{align}
where momentum and energy are
\begin{equation}\label{eq:n=0energy}
  k_l=\frac{2\pi l}{L}
  \;,\;\;
  l\in\mathbf{Z}
  \;,\quad
  E_l=\sqrt{F^2+k_l^2}
  \;.
\end{equation}
Note that for all nonzero momenta there are two degenerate states
with equal energy, corresponding to left and right moving particles
(and right and left moving antiparticles with negative energy).  The
state with $k=0$, $E=F$ is not degenerate.

In the case of $n=1$ bosonic vacuum (with $A_1=\frac{2\pi}{eL}$, $A_0=0$, and
$\phi=v\e^{i\frac{2\pi x}{L}}$) and periodic boundary conditions%
\footnote{Alternatively one could use the equations in trivial
  background and impose anti-periodic boundary conditions.}
we get
\begin{align}\label{eq:psiplusn=1}
  \Psi_{+} &=
    \e^{-iE_lt}\begin{pmatrix}
      -\e^{i\frac{2\pi l}{L}x}F \\
      \e^{i\frac{2\pi (l-1)}{L}x}(E_l-k_l)
	       \end{pmatrix}
  \;,\\
  \Psi_{-} &= \hphantom{\scriptstyle -}
    \e^{iE_lt}\begin{pmatrix}
      \e^{i\frac{2\pi l}{L}x}(E_l-k_l) \\
      \e^{i\frac{2\pi (l-1)}{L}x}F
	      \end{pmatrix}
  \;,\notag
\end{align}
with momenta and energy
\begin{equation}\label{eq:n=1energy}
  k_l =\frac{2\pi (l-\frac{1}{2})}{L}
  \;,\;\;
  l\in\mathbf{Z}
  \;,\quad
  E_l =\sqrt{F^2+k^{2}}
  \;.
\end{equation}
There is no state with $k=0$ in this case, and all states are doubly
degenerate in energy.

We see, that the fermion spectra in bosonic vacua with even and odd
topological numbers are indeed different.  So, in case of finite
space size, a gauge transformation with odd $n$ leads to physical
changes in the system.  We thus should say that the only allowed
gauge transformations (i.e.\ those that connect physically
indistinguishable field configurations) have even
$n=\frac{1}{2\pi}(\alpha(L)-\alpha(0))$.  Transitions between states
with bosonic background being vacuum configurations with $n=0$ and
$n=1$ are still possible, but they are just tunneling between
different (local) minima of the energy of the system (see
Fig.~\ref{vacua}).

In the limit of infinite space ($L\to\infty$), however, the difference
between energy levels disappears.  The total vacuum energy (or Dirac
see energy) also turns out to be equal in both $n=0$ and $n=1$
backgrounds in infinite space limit, see Appendix~\ref{sec:ave}.
Calculation of the fermion number of the Dirac see in these
backgrounds, performed in Appendix~\ref{sec:appFnumber}, gives zero in
both backgrounds.  In the limit of infinite space transitions from
$n=0$ to $n=1$ are again vacuum to vacuum transitions, while the vacua
are not exactly gauge equivalent, but rather simply degenerate.

\subsection{Level crossing picture}

\begin{figure}
  \begin{center}
    \includegraphics[width=\columnwidth]{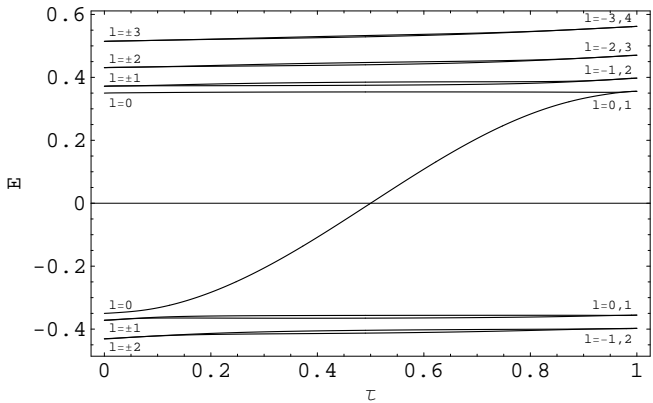} 
    \caption{Fermionic energy levels in the background (\ref{spht})
      obtained numerically for finite space of length $L=50$
      and periodic boundary conditions. 
      Fermion mass $F=0.35$,
      the charge $e=1$.}
    \label{flc} 
  \end{center} 
\end{figure}

Let us analyze a process in external gauge and Higgs fields
interpolating between adjacent bosonic vacua, for example
\begin{align}\label{spht}
  \phi ^{cl}(x,\tau) &=
    \frac{v}{\sqrt{2}}\e^{-\frac{2\pi ix\tau }{L}}\big[ \cos (\pi
    \tau ) \notag\\
    &\qquad+i\sin (\pi \tau )\tanh (m_H x\sin (\pi \tau )\big]
  \;, \notag\\
  A_{1}^{cl}(x,\tau) &=
    -\frac{2\pi \tau }{eL}
  \;,
\end{align}
with parameter $0<\tau<1$.  This configuration goes from the vacuum
$n=0$ at $\tau=0$ to $n=-1$ at $\tau=1$ minimizing the energy of the
intermediate configurations \cite{Bochkarev:1987wg}.
For each value of the parameter $\tau$ we solved numerically the
static Dirac equation
$H_{D,\tau}\Psi_\tau =E_\tau\Psi_\tau$.  Evolution of the energy
levels is presented in Figure~\ref{flc}.  Exactly one level (level
with negative energy with $l=0$ in (\ref{n=0})) crosses zero.
Together with the positive energy level with $l=0$ they merge into the
two degenerate energy states with $l=0$ and $l=1$ in $n=-1$ vacua (see
(\ref{eq:psiplusn=1}), or, to be more precise, they go to linear
combinations of the $l=0$ and $l=1$ states in (\ref{eq:psiplusn=1})).  

So exactly one fermion should be created in a process with gauge
fields interpolating between $n=0$ and $n=-1$ bosonic vacua.

\section{Instanton calculation of the cross sections}
\label{sec:instantons}

The level crossing picture described in the previous section does not
allow to calculate the probabilities of real processes of one fermion
creation (or decay) at low energies.  Convenient method for the
calculation of such probabilities is given by perturbation theory in
the instanton background
\cite{'tHooft:1976fv,Krasnikov:1979bh,
Krasnikov:1978dg,Ringwald:1990ee,Espinosa:1990qn}.

The usual prescription is to calculate the Euclidean Greens' functions
in instanton background and then apply the LSZ reduction procedure to
get matrix elements.  The fermionic part of the Green's function
contains the fermionic determinant in the instanton background
calculated without the zero mode.
However, the determinant of the Dirac operator $K$ for a chiral
fermion in nontrivial background is hard to define.
The operator $K$ itself maps from a Hilbert space to another and 
its determinant is not defined.  The usual trick is to use instead
$K^\dagger K$ or $KK^\dagger$.  However in non-trivial background, 
these two operators do not contain the same number of zero modes. 
Their determinants, after removing the relevant zero-mode still differ
by a constant.

This problem
seems to be connected with the fact that usual normalization is
performed by division by the vacuum partition function\footnote{More
  precisely by the
  determinant of the Dirac operator in the trivial background.} while
the Hilbert spaces for fermionic wave functions are not exactly the
same in trivial and one instanton backgrounds \cite{Nielsen:1976hs}.
We have to emphasize
that this subtlety is not a feature of 1+1--dimensional models but
is present in the Standard Model also. In existing calculations of
chiral fermion contribution to the instanton transitions, the
corresponding normalization was defined using dimensional arguments
only \cite{Espinosa:1990qn,Vainshtein:1981wh}. We propose the definition 
of the required determinant using sort of a
valley approximation for the path integral.

In this section we describe the whole procedure in detail.  In
subsection~\ref{sec:sub4A} we describe the instanton solution and the
zero modes.  Subsection~\ref{sec:eucl-greens-funct} is devoted to the
na\"ive definition of the Euclidean Greens' functions (and the
fermionic determinant) which leads to an inconsistent result.  
In the subsection~\ref{sec:Detdefin} we
describe a careful definition of the fermionic determinant that
resolves the problem.  In the last subsection~\ref{sec:LSZ} the LSZ
reduction formula is used to get matrix elements.

\subsection{Instanton solution and fermionic zero modes}
\label{sec:sub4A}

Let us first review the Euclidean formulation of the model we use.
It is described in more detail in Ref.~\cite{Burnier:2005he}
 
The Lagrangian (\ref{Lm}) may be rewritten in Euclidean space: 
\begin{equation}\label{L}
  \mathcal{L}^{E} = \frac{1}{4}F_{\mu \nu }F_{\mu\nu}
  +\frac{1}{2}(D_{\mu }\phi )^{\dagger }(D_{\mu }\phi)
  +V(\phi )
  +\overline{\Psi}K\Psi
  \;,
\end{equation}
with
\begin{equation}\label{hatD}
  {K} = i\gamma^E_\mu\Dm
    -if\left(\Pleft\phi^*-\Pright\phi\right)
  \;,
\end{equation}
and $\gamma_0^E=i\gamma^0$, $\gamma_1^E=\gamma^1$.  The fields
$\overline{\Psi }$ and $\Psi$ are independent variables in Euclidean
case, and the gauge transformation reads:
\begin{eqnarray} 
  \Psi  &\longrightarrow &e^{i\alpha (x)\frac{\gamma _{5}}{2}}\Psi 
    \quad,\;
  \overline{\Psi}  \longrightarrow 
    \overline{\Psi }e^{i\alpha (x)\frac{\gamma _{5}}{2}}
  \;,\notag \\ 
  \phi  &\longrightarrow &e^{i\alpha (x)}\phi
  \;. \label{eucgauge}
\end{eqnarray}
For comparison, the Lorentz transformation is: 
\begin{eqnarray*} 
  \Psi(x)  &\rightarrow & \Psi'(x')= \Lambda_s \Psi(\Lambda^{-1} x')
  \;, \\
  \overline{\Psi }(x) &\rightarrow 
  &\overline{\Psi}'(x')=\overline{\Psi }\Lambda_s ^{-1}(\Lambda^{-1} 
  x')
  \;,
\end{eqnarray*} 
with $\Lambda_s=\exp(i\gamma^5\frac{\theta}{2})$ being the spinor
rotation matrix in two dimensions.

\paragraph{Instanton solution.}
 
The instanton describing the tunneling between the states
$|0\rangle$ and $|n\rangle$ is simply the Nielsen--Olesen vortex with
winding number $n$~\cite{Nielsen:1973cs}, which is a solution of the
Euclidean equations of motion derived from the Lagrangian~(\ref{L}).
It is obtained  by using the following Anzats, which is the most
general Anzats consistent with symmetry under spatial rotations
accompanied by the corresponding gauge transformations,
\begin{eqnarray} 
  \phi (r,\theta ) &=& \e^{in\theta }\phi(r) \equiv \e^{in\theta }vf(r)  
  \;,\label{phi} \\
  A^{i}(r,\theta ) &=&\varepsilon 
  ^{ij}\widehat{r}^{j}A(r)   
  \;,\label{A}
\end{eqnarray}
where $\widehat{r}=(\cos\theta,\sin\theta)$ is the unit vector
and $\varepsilon^{ij}$ is the
completely antisymmetric tensor with $\varepsilon ^{01}=1$.  The
functions $A$ and $f$ have to satisfy the following limits:
\begin{align*}
  f(r)&\overset{r\rightarrow 0}{\longrightarrow }c r^{\left| n\right| } 
  \;,&
  f(r)&\overset{r\rightarrow \infty }{\longrightarrow }1
  \;,\notag \\ 
  A(r)&\overset{r\rightarrow 0}{\longrightarrow }0
  \;,&
  A(r)&\overset{r\rightarrow \infty }{\longrightarrow }-\frac{n}{er}
  \;.
\end{align*}
The number $\Delta N$ of fermions created in the instanton transition 
can be computed by integrating (\ref{curentanom}) over the Euclidean 
space
\begin{equation*}
  \Delta N=-\int d^2x\partial_\mu 
  J_\mu=-\int d^2x \frac{e}{4\pi}\varepsilon_{\mu\nu}F_{\mu\nu}  
  =-q
  \;,
\end{equation*}
where $q=\int d^2x \frac{e}{4\pi}\varepsilon_{\mu\nu}F_{\mu\nu}$ is
the winding number of the gauge field configuration.  For the
instanton configuration (\ref{phi},\ref{A}), we have $q=n$.

At large $r$ both approach their asymptotic exponentially
\begin{align*}
  f(r) &\overset{r\to\infty}{\longrightarrow}
    1-f_0 \sqrt{\frac{\pi}{2r}}\e^{-mr}
  \;, \\
  A(r) &\overset{r\to\infty}{\longrightarrow}
    -\frac{n}{er}+\frac{a_0}{e}\sqrt{\frac{\pi}{2r}}\e^{-m_Wr}
  \;.
\end{align*}

Later we will also use this solution in unitary gauge, i.e.\ gauge where
$\phi(r,\theta)\overset{r\to\infty}{\longrightarrow}v$ for all
directions.  The solution in this gauge is singular at the origin, but
the singularity is a gauge artefact.  Also, for odd $n$, the
fermion zero mode (see next paragraph) is not a single valued function in
unitary gauge.  However, one may also think of configuration in
unitary gauge as a limit of the configuration (\ref{phi},\ref{A}) 
transformed with the gauge function
\begin{equation*}
  \alpha(\theta)=-n(\theta-2\pi\Theta^\epsilon(\theta-\pi))
  \;,
\end{equation*}
where $\Theta^\epsilon$ is a function approaching the step function
for vanishing $\epsilon$.

\paragraph{Fermionic zero modes.}

According to the index theorem (see for example~\cite{Bertlmann}), the
Dirac operator in the background of the instanton satisfies the
following relation: $\dim \ker[K]-\dim\ker[K^\dagger]=n$.  As the
instanton in $1+1$ dimensions coincides with the vortex, these zero
modes may be found by carrying out a similar analysis as in
\cite{Jackiw:1981ee}; where the fermionic zero modes on the
Nielsen--Olesen string were analyzed for non chiral fermions.  In this
subsection we present the corresponding equations.
 
 
The zero modes are the regular normalizable solutions of the equation 
$K \Psi=0$, with $A_\mu$ and $\phi$ given by (\ref{phi},\ref{A}).
Using spherical mode expansion of the form
$\Psi(r,\theta)=\exp\left[-\int_{0}^{r} 
\frac{A(\rho)}{2}d\rho\right]\sum_{m=-\infty}^{\infty}e^{im\theta}\psi^m(r)
$ we get
\begin{eqnarray*}
  Ff(r)\psi^m_L-\left(\frac{\partial}{\partial r} 
    - \frac{m-n-1}{r}\right)\psi_R^{m-n-1}
  &=& 0
  \;,\notag\\ 
  \left(\frac{\partial}{\partial r}
  +\frac{m}{r}\right)\psi_L^{m}-Ff(r)\psi_R^{m-n-1}
  &=&0
  \;,
\end{eqnarray*}
where $F=fv$ is the fermion mass.  We also continue to use indices $L$ and
$R$ to denote two components of the spinor, though they
are no longer left and right moving in Euclid.  In our case, the
analysis of \cite{Jackiw:1981ee} shows that for a vortex with topological
number $n<0$ there are exactly $|n|$ fermionic zero modes in the
spectrum of $K$ with $m$ in the interval $m\in \{-n+1,..,1,0\}$ and
none in the spectrum of $K^\dagger$.  For $n>0$ there are no zero
modes in the spectrum of $K$, but $n$ in the spectrum of $K^\dagger$.

For the case of $n=-1$ studied in~\cite{Burnier:2005he} the explicit form
of the zero mode is given by
\begin{align*}
  \Psi^0_L(r) &=-\Psi^0_R(r) \notag\\
    & = \const \cdot \exp\left( -\int_0^r \left\{Ff(r')+
      \frac{e}{2}A(r')\right\}dr'\right) \notag\\
    & \overset{r\to\infty}{\longrightarrow}
      U_0\frac{\e^{-F r}}{\sqrt{r}}
  \;.
\end{align*}
Note that for massless fermions ($F=0$), the zero mode decreases as
$\frac{1}{\sqrt{r}}$ for large $r$. It is therefore not normalizable
and has a divergent action.

\subsection{Euclidean Greens functions}
\label{sec:eucl-greens-funct}

Let us start from evaluating the generating functional for fermionic
Euclidean Green's functions.  We will not write here the source terms
for bosonic fields explicitly because there is no problem of dealing
with the bosonic part here, see, eg.~\cite{Baacke:1994bk}).
\begin{align}
  Z[\bar\eta,\eta] &=
    \frac{1}{Z_0}\int\! \cD A_\mu \cD\phi\: \e^{-S_\mathrm{bosonic}}
    Z_{A,\phi}[\bar\eta,\eta]
  \;,\notag\\
  Z_{A,\phi}[\bar\eta,\eta] &= \int\! \cD\Psi \cD\overline{\Psi}\,
    \e^{-\!\int\! d^2x \, (\overline{\Psi}{K}\Psi
      -\bar\eta\Psi-\overline{\Psi}\eta
      )}
  \;. \label{eq:Zpsipsi}
\end{align}
where $Z_0$ is the same functional integral with zero source terms.
At one-loop level the fermionic part of the generating functional can
be calculated regarding the bosonic fields $A_\mu$, $\phi$ as external
classical sources, both in the generating functional itself and in the
normalization factor $Z_0$, which then factorizes in bosonic and
fermionic parts.

Let us try to evaluate the fermionic part
$Z_{A,\phi}[\bar\eta,\eta]/Z_0$.  As far as it is just a Gauss
integral over Grassman variables we can (at least formally) perform it
exactly.  To define it we proceed in the spirit of
Ref.~\cite{Krasnikov:1979bh,Krasnikov:1978dg}.

Let us start with the trivial background case first.  We define the
following eigenvalues and eigenvectors
\begin{equation}\label{vacDeigensystem}
  {K_0^\dagger}{K_0} \rho_n = \kappa^2_n\rho_n
  \;,\quad
  {K_0}{K_0^\dagger} \tilde\rho_n = \kappa^2_n\tilde\rho_n
  \;,
\end{equation}
where $K_0$ is the Dirac operator~(\ref{hatD}) in zero
background, and the eigenvectors $\tilde\rho$ and $\rho$ are
normalized to 1 and connected with the formula
\begin{equation}\label{vacrhotilde}
  \tilde\rho_n = \frac{1}{\kappa_n}{K_0}\rho_n
  \;.
\end{equation}
Several notes are required here.  First, the operators
${K_0}{K_0^\dagger}$ and ${K_0^\dagger}{K_0}$ are self conjugate, and
thus the sets $\rho_n$ and $\tilde{\rho}_n$ form full orthonormal sets
of functions.  Second, we are not trying to use operators ${K}$ (or
${K^\dagger}$) to define the eigenfunctions because they map from the
space of spinors $\Psi$ to a space with different gauge transformation
properties (see~(\ref{eucgauge})).  And finally, as far as the
background is now just the trivial vacuum, all $\kappa_n\neq0$, so the
relation~(\ref{vacrhotilde}) holds for all $n$.  Also, by convention,
we choose all $\kappa_n>0$.

Now we expand fermionic fields using these eigenmodes
\[
  \Psi = \sum_n a_n\rho_n
  \;,\quad
  \overline{\Psi} = \sum_n \bar{a}_n\tilde{\rho}_n^\dagger
\]
and define the functional integral measure as
\[
  D\Psi D\overline{\Psi} = \prod_n da_nd\bar{a}_n
  \;.
\]
Then the integration immediately leads to
\begin{align*}
  Z_0 &= \int\! \cD\Psi \cD\overline{\Psi}\,
      \exp\left[-\int d^2x \overline{\Psi}{K_0}\Psi \right]\\
  & = \int \prod_n da_nd\bar{a}_n
      \exp\left[-\sum_n\kappa_n \bar{a}_na_n\right]
    = \prod_n \kappa_n
  \;.
\end{align*}
Analogous procedure we should also apply in the nontrivial background.
We find the eigenvalues of the two following equations
\begin{equation}\label{instDeigensystem}
  {K^\dagger}{K}\psi_n = \lambda_n^2\psi_n
  \;,\quad
  {K}{K^\dagger}\tilde{\psi}_n = \lambda_n^2\tilde{\psi}_n
  \;,
\end{equation}
with relation similar to~(\ref{vacrhotilde}) for all $\lambda_n\neq0$
\begin{equation}\label{psitilde}
  \tilde\psi_n = \frac{1}{\lambda_n}{K}\psi_n
  \;.
\end{equation}
In nontrivial background there may also exist zero eigenvalues, and
${K}$ is no longer a normal operator\footnote{Normal operator is an
operator $A$ with the property $A^\dagger A= AA^\dagger$.}, so there
may be different number of zero eigenvalues for ${K^\dagger}{K}$ and
${K}{K^\dagger}$.  The index theorem says that
$\dim\Ker{K^\dagger}{K}-\dim\Ker{K}{K^\dagger}=n$, so in one instanton
case there should be one more zero mode for ${K^\dagger}{K}$ (and it
is the only zero mode present).  For zero modes there is no relation
of the type~(\ref{psitilde}), and we simply define them as
\begin{equation*}
  {K^\dagger}{K}\psi^0_k = 0
  \;,\quad
  {K}{K^\dagger}\tilde{\psi}^0_l = 0
  \;,\quad\mathrm{with } \int|\psi^0|^2d^2x<\infty
  \;.
\end{equation*}
Now we re-expand fermionic fields in terms of the new orthonormal sets
$\psi_K=\{\psi^0_k,\psi_n\}$ and
$\tilde{\psi}_L=\{\tilde{\psi}^0_l,\tilde{\psi}_n\}$
\[
  \Psi = \sum_k c_k\psi^0_k+\sum_n b_n\psi_n
  \;,\quad
  \overline{\Psi} = \sum_l \bar{c}_l\tilde{\psi}^{0\dagger}_l
    +\sum_n \bar{b}_n\tilde{\psi}^\dagger_n
  \;.
\]
One should now take care when defining the integration measure, to be
consistent with~(\ref{vacDeigensystem})
\[
  D\Psi D\overline{\Psi} =
    P\prod_k dc_k \prod_l d\bar{c}_l \prod_n d b_n d\bar{b}_n
  \;,
\]
where $P$ is the Jacobian for the change of variables
$\{a_n,\bar{a}_n\}\to\{c_k,b_n,\bar{c}_l,\bar{b}_n\}$
\[
  P[A,\phi]=\det[(\rho_n,\psi_k)]^{-1}
    \det[(\tilde{\psi}_l,\tilde{\rho}_n)]^{-1}
  \;,
\]
where $(\alpha,\beta)=\int dx\bar\alpha(x)\beta(x)$ denotes scalar
product for spinor functions.  Absolute value of $P$ is one, because
it corresponds to transition between full orthonormal sets of
functions, so it is only a complex phase, which, in general, depends
on the background fields $A_\mu$,~$\phi$.  As noted
in~\cite{Krasnikov:1979bh,Krasnikov:1978dg} it is essential to take
this phase into account to reconstruct correct perturbative expansion
for the theory.  In our case, in the leading one-loop approximation
this is not important, because there are no instanton orientation to
be integrated over---instanton field configurations differ only by
translations and gauge transformation.  Note that for example in four
dimensional nonabelian theory this is not the case.

Performing Gaussian integration over $dc_kd\bar{c}_ldb_nd\bar{b}_n$ in
(\ref{eq:Zpsipsi}) we get
\begin{align}\label{Zaphi_result}
  Z_{A,\phi}[\bar{\eta},\eta] = P[A,\phi]
    &\times \prod_n\left(\lambda_n
       +(\bar{\eta}^\dagger,\psi_n)(\tilde{\psi}_n,\eta)
     \right) \notag\\
    &\times
      \prod_k(\bar{\eta}^\dagger,\psi^0_k)\times
      \prod_l(\tilde{\psi}^0_l,\eta)
  \;.
\end{align}
This formula leads to the standard result that nonzero Greens'
functions must contain in addition to usual even number of fermionic
legs a set of fermionic operators of a special structure, defined by
fermionic zero modes.  In the instanton case we have only one zero
mode, and the simplest nonzero Green's function is given formally by
the following expression
\begin{multline}\label{eq:Zdetprime}
  \left.\left[\frac{1}{Z_0}
    \frac{\delta Z_{A,\phi}[\eta,\bar{\eta}]}{\delta\bar{\eta}}
  \right]\right|_{\eta,\bar{\eta}=0} 
  =
  \left[\frac{\prod_{n\neq0}\lambda_n}{\prod_n\kappa_n}\right] \times
    P[A,\phi] \times \psi_0
  \\
  \equiv \sqrt{{\det}_\mathrm{ren}[K_I^\dagger K_I]} \times
    P[A,\phi] \times \psi_0
  \;.
\end{multline}
It is easy to see that this quantity is ill defined.  The left hand
part of the equality has dimension $m^{1/2}$.  In the right hand part
of the expression $\psi_0$ has dimension $m$ (as it is normalized to
one), $P$ is dimensionless.  Thus, the dimension of the infinite
product should be $m^{-1/2}$, and not $m^{-1}$, as could be expected
na\"ively.

\subsection{Determinant definition}
\label{sec:Detdefin}

Let us try to clarify the definition of the determinant.  The problem
with the description in the previous section is that, strictly
speaking, the eigensystems in (\ref{vacDeigensystem}) and in
(\ref{instDeigensystem}) generally belong to different Hilbert
spaces---fermions living in trivial and one instanton backgrounds.
One may hope that the situation can be cured if one calculates a
quantity in a trivial background.  A good candidate is the expectation
value for two fermion operators in external instanton--antiinstanton
background
\begin{multline}\label{eq:AIpsipsi}
  \bra{0}\overline\Psi(T) \Psi(-T)\ket{0}_{I-A}= \\
  \int\! \cD\Psi \cD\overline{\Psi}\,
    \exp\left[-\int d^2x (\overline{\Psi}{K}_{I-A}\Psi)\right]
    \overline\Psi(T)\Psi(-T)
  \;,\raisetag{9ex}
\end{multline}
where index $I-A$ means that everything is calculated in the
instanton--antiinstanton background, with instanton and antiinstanton
centered at Euclidean time $t_0$ and $-t_0$ respectively.  Just by
construction for large $t_0$ this reproduces the modulus squared of
the one fermion expectation value in instanton background
\begin{multline}\label{eq:det2}
  \bra{0}\overline\Psi(t_0+T) \Psi(-t_0-T)\ket{0}_{I-A}
  \to |\langle|\Psi(-T)|\rangle_I|^2 \\
  \text{ for } t_0\to\infty
  \;.
\end{multline}
Let us now calculate this integral using the method described in
Section~\ref{sec:eucl-greens-funct}.  We get the eigensystems of the
form
\begin{align}\label{eq:DIAeigensystem}
  K_{I-A}^\dagger K_{I-A} \Psi_N &= \Lambda^2_N\Psi_N
  \;, \\
  K_{I-A}K_{I-A}^\dagger \tilde\Psi_N &=
    \Lambda^2_N\tilde\Psi_N
  \;, \notag
\end{align}
where now there are no exact zero modes for both operators, so all
eigenfunctions are related by a relation of the form
(\ref{psitilde}).  However, we can immediately construct an
approximate eigensystem for (\ref{eq:DIAeigensystem})
\begin{equation*}
  \begin{array}{c@{\,=\{\,}ccc@{\,\}\;}l}
    \Lambda_N    & \lambda^I_n;             & \lambda^A_n;
                   & \Lambda_0             & ,\\
    \Psi_N       & \psi^I_n(t-t_0);       & \psi^A_n(t+t_0);
                   & \psi^I_0(t-t_0)       & ,\\
    \tilde\Psi_N & \tilde\psi^I_n(t-t_0); & \tilde\psi^A_n(t+t_0);
                   & \tilde\psi^A_0(t+t_0) & ,
  \end{array}
\end{equation*}
where $\Lambda_0$ is small and goes to zero as $t_0\to\infty$.  So
there are two sets of modes, corresponding to nonzero eigenmodes of
the instanton and antiinstanton centered at their locations, and one
nearly zero mode $\Lambda_0$, which is constructed out of a zero mode
for instanton for $\Psi$ and for antiinstanton for $\tilde\Psi$.

It is now trivial to calculate (\ref{eq:AIpsipsi}) using
(\ref{Zaphi_result}) and differentiating it by
$\delta\eta\delta\bar\eta$
\begin{equation*}
  \bra{0}\overline\Psi(T) \Psi(-T)\ket{0}_{I-A} \!=\!
    \frac{1}{Z_0}
    (\prod_N \Lambda_N )\sum_N\frac{\Psi_N(-T)\overline{\tilde\Psi}_N(T)}{\Lambda_N}
  .
\end{equation*}
The sum is governed by the term with $\Lambda_0$, so we get
\begin{equation}\label{eq:detdetprime}
  \bra{0}\overline\Psi(T) \Psi(-T)\ket{0}_{I-A} =
    \frac{(\prod_n \lambda^I_n)}{(\prod_n \kappa_n)}
    \frac{(\prod_n \lambda^A_n)}{(\prod_n \kappa_n)}
    \Psi_0(-T)\overline{\tilde\Psi}_0(T)
\end{equation}
(no zero mode is present in $\prod_n \lambda^I_n$).  It is easy to
see, comparing formulas (\ref{eq:Zdetprime}), (\ref{eq:detdetprime})
and (\ref{eq:det2}) that
\begin{multline}\label{eq:detcorrect}
  \langle|\Psi(-T)|\rangle_I =
    \sqrt[4]{
      \frac{{\det}'[K_I^\dagger K_I]}{{\det}[K_0^\dagger K_0]}
      \frac{\det[K_A^\dagger K_A]}{\det[K_0^\dagger K_0]}
    }
    \;\psi^I_0(-T) \\
    \equiv  \sqrt{{\det}_\mathrm{ren}[K_I^\dagger K_I]} \times
    \psi_0
  \;,
\end{multline}
up to some complex phase, in principle.  Calculation and
renormalization of the determinant ${\det}'[K_I^\dagger K_I]$ is
described in detail in Ref.~\cite{Burnier:2005he} and additional
subtleties for calculation of the antiinstanton determinant, which has
no zero mode, is given in Appendix~\ref{sec:detcalculation}.  We can
then use~(\ref{eq:detcorrect}) as the correct definition of the
renormalized determinant in the one instanton background.  The
dimension of the ratio $\frac{{\det}'[K_I^\dagger
K_I]}{{\det}[K_0^\dagger K_0]}$ is $m^{-2}$ (zero mode is absent in
the numerator), $\frac{{\det}'[K_A^\dagger K_A]}{{\det}[K_0^\dagger
K_0]}$ has dimension zero (no zero mode here), and $\psi_0$ is $m$
because of normalization.  This whole expression has dimension
$m^{1/2}$, which is now correct.

\subsection{Reduction formula}
\label{sec:LSZ}

A convenient method to get physical amplitudes from the Greens'
functions is provided by LSZ reduction procedure.  There is one
subtlety in application of the reduction formula in the instanton
case, as compared to usually considered topologically trivial
situations.  The reduction formula is derived using the assumption
that field operators are connected with creation-annihilation
operators of the physical particles in the same canonical way for all
times (both initial and final).  For instanton like configurations
this is true only in unitary gauge, which is singular at the origin.
However, this singularity is of purely gauge type and does not
contribute to the poles of the Green's function, so it is safe to use
it.  At the same time other gauge choices may lead to appearance of
nonphysical singularities in the Green's function.

We start from the Euclidean Green's function, calculated in the saddle
point approximation
\begin{multline*}
  \langle \Psi(x) h(y_1) \dots h(y_m) \rangle_\mathrm{inst}
  = \\
    \int \! d^2x_0 \, J(\langle\phi\rangle) \det[K_\mathrm{scalar}]^{-1/2}
    \sqrt{{\det}_\mathrm{ren}[K_I^\dagger K_I]} \e^{-S_\mathrm{inst}} \\
    \times
    \psi_0(x-x_0)
    h_\mathrm{inst}(y_1-x_0)\dots h_\mathrm{inst}(y_m-x_0)
  \;,
\end{multline*}
where $\det[K_\mathrm{scalar}]$ is the determinant of the bosonic
field quadratic excitations over the instanton background, see
eg.~\cite{Baacke:1994bk}, $J(\langle\phi\rangle)$ is the Jacobian
appearing from the transition to the integration over the collective
coordinate $x_0$---instanton center,
$\det_\mathrm{ren}[K^\dagger_IK_I]$ is the fermionic determinant
defined in the previous subsection, $\psi_0$ is the fermionic zero
mode, and $h_\mathrm{inst}=\phi_\mathrm{inst}-\phi_v$ is the instanton
solution for the deviation of the scalar field from vacuum value.  In
complete analogy it is possible to add gauge fields here.  Also pairs
of fermion fields can be added, connected with fermion propagator in
instanton background.

The meaning of integration over the position of the instanton is clear
after going to the momentum representation, where it leads to the
energy-momentum conservation
\begin{multline*}
  (2\pi)^2\delta^2(p+k_1+\dots+k_m) \tilde{G}(p,\{q\}) =\\
  \int \!d^2x\,d^2y_1\dots d^2y_m
  \e^{ipx}\e^{ik_1y_1}\dots\e^{ik_my_m} \\
   \times\langle \Psi(x) h(y_1) \dots h(y_m) \rangle_\mathrm{inst}
  \;.
\end{multline*}
Using these formulas we get for the Green's function in momentum
representation
\begin{multline}\label{Gpq}
  \tilde{G}(p,\{q\}) = J(\langle\phi\rangle) \det[K_\mathrm{scalar}]^{-1/2}
  \left({\det}_\mathrm{ren}[K_I^\dagger K_I]\right)^{1/2} \\
  \times\e^{-S_\mathrm{inst}} \times
  \psi_0(p)
  h_\mathrm{inst}(k_1)\dots h_\mathrm{inst}(k_m)
  \;,
\end{multline}
where $\psi_0(p)$, $h_\mathrm{inst}(k)$ are the Fourier transforms of
the zero mode and the instanton respectively,
\[
  \psi_0(p) = \int \!d^2x\, \e^{ipx} \psi_0(x)
  \;,
\]
etc.

\paragraph{Fourier transforms.}

Let us now calculate Fourier transforms appearing in (\ref{Gpq}).  To
get the matrix elements we will be interested only in the pole terms
at the physical mass, so we can analyze only infinite contributions
from the exponential tails of the solutions.

The instanton solution for the scalar field is (see.~\cite{Burnier:2005he})
\[
  h_\mathrm{inst}(x) = v(1-f(r)) \simeq v f_0 K_0(m_Hr) \;,
\]
where the constant $f_0$ is determined from the asymptotics of the
exact solution $1-f(r)$ at large $r$ ($r$ is the distance from the
instanton origin in Euclid).  Thus we get
\[
  h_\mathrm{inst}(k) = \int \!d^2x\, \e^{ikx} h(x)
  = -\frac{2\pi f_0v}{m_H^2+k^2}+\text{regular terms}
  \;.
\]
For the fermion zero mode we have
\[
  \psi_0(x) = \begin{pmatrix} \psi_{0L} \\ \psi_{0R} \end{pmatrix}
  \to_{r\to\infty}
  \begin{pmatrix} \e^{-i\theta/2} \\- \e^{i\theta/2} \end{pmatrix}
    U_0 \frac{\e^{-F r}}{\sqrt{r}}
  \;,
\]
where the constant $U_0$ is defined from the exact numerical solution 
for the zero
mode and normalization $\int \psi^\dagger_0\psi_0 d^2x=1$.  The
function $\psi_0(x)$ is not well defined in singular gauge, as far 
as it changes
sign when $\theta$ changes by $2\pi$.  We can say that $\theta$ runs
from $-\pi$ to $\pi$ only, i.e.\ put the cut along the negative $x$
(space coordinate) axis%
\footnote{The singular gauge can be considered as a limit of gauges
  obtained by applying smooth gauge transformation with gauge function
  $\alpha=\theta+2\pi\Theta^\epsilon(\theta-\pi)$ to the instanton
  solution, with $\Theta^\epsilon$ being a smooth function becoming
  the step function in the limit $\epsilon\to0$.}.
It is simpler in this case to make calculations after setting
explicitly $k_1=0$, then we get for the Fourier transform (in
Minkowski)
\begin{multline*}
  \psi_{0R,L}(k_0) = \\
  \mp U_0\sqrt{2\pi}\frac{\sqrt{k_0\pm k_1}}{F}
  \left(
    \frac{\e^{\mp i\pi/4}}{F-\sqrt{k_\mu k_\mu}}+
    \frac{\e^{\pm i\pi/4}}{F+\sqrt{k_\mu k_\mu}}
  \right) \\
  +\text{regular terms}
  \;,
\end{multline*}
where upper and lower signs correspond to $\psi_{0R}$ and $\psi_{0L}$
respectively.

\paragraph{Matrix element.}

As an example let us calculate the matrix element with one fermion and
two scalars.  It is given by (in Minkowski space-time)
\begin{multline*}
  iM(p,k_1,k_2) = i\bar{v}(p)(\hat{p}+F) \psi_0(p)
  \times \\
  (-i)(k_1^2-m_H^2) h_\mathrm{inst}(k_1)
  \times 
  (-i)(k_2^2-m_H^2) h_\mathrm{inst}(k_2)
  \times \\
  J \det[K_\mathrm{scalar}]^{-1/2}
  \sqrt{{\det}_\mathrm{ren}[K_I^\dagger K_I]}
  \e^{-S_\mathrm{inst}}
  \;.
\end{multline*}
Here $\bar{v}(p)$ is the antifermion spinor normalized like
$v(p)\bar{v}(p)=\hat{p}-m$.  So, the matrix element is
\begin{multline}\label{mat}
  i M(p,k_1,k_2) = i\sqrt{4\pi} U_0 (2\pi f_0v)^2
    J \times\\
    \det[K_\mathrm{scalar}]^{-1/2}
    \left({\det}_\mathrm{ren}[K_I^\dagger K_I]\right)^{1/2}
    \e^{-S_\mathrm{inst}}
  \;.
\end{multline}
We get a non-zero Lorentz invariant matrix element for a process
involving one fermion and two bosons, as announced previously. 

The matrix element (\ref{mat}) arise for instance in processes where
an antifermion $\overline{\Psi}$ decays into two scalar $\phi$ if
$F>2m_H$.  One may also analyze other Greens' functions.  For
instance, even simpler Green's function of the form $\langle \Psi
h\rangle_\mathrm{inst}$ is nonzero in the model, giving boson-fermion
mixing.

\section{Conclusions}
\label{sec:concl}

We have analyzed the Abelian Higgs model in 1+1 dimensions.  Half
charged chiral fermions with mass generated by Higgs mechanism in
this model are created in processes which change the topological
number of the vacuum.  A peculiar feature of the 1+1 dimensional
models makes it possible to create only one fermion in the process where
topological vacuum number changes by one.  Unlike in similar 3+1
dimensional models, this model does not possess Witten anomaly.
Neither this effect contradicts Lorentz symmetry in 1+1 dimensions.

We calculated the probability of such process using perturbation
theory in instanton background.  Calculation of this probability
requires evaluation of the fermionic determinant in one instanton
background.  We note (see Section~\ref{sec:Detdefin}) that the
fermionic determinant for chiral fermions is very hard to define in
topologically nontrivial background, with the main obstacle lying in
the correct normalization, which usually requires division by fermion
determinant in zero (topologically trivial) background.  We want to
emphasize, that this problem arises exactly in the same form in 3+1
dimensional theories (separately for each fermionic doublet in case of
SU(2) theory).  Up to our knowledge the relevant normalization was
chosen only on dimensional grounds in
literature~\cite{Espinosa:1990qn,Vainshtein:1981wh}.  We propose a
method to deal with the problem in 1+1 dimensions, though direct
generalization of it to more dimensions is not trivial.

The authors are grateful to V.~Rubakov, P.~Tinyakov, S.~Dubovsky,
S.~Khlebnikov for helpful discussions on the subject.  The work of
F.B. was supported in part by INTAS YSF 03-55-2201 and Russian Science
Support Foundation.  The work of Y.B. and M.S. is supported by the Swiss
Science Foundation.

\appendix
\section{Vacuum energy}\label{sec:ave}

Let us calculate Dirac sea energy in the bosonic vacua with odd and even
topological charges.

In sector with $n=0$ the Dirac sea energy in a box of size $L$ is
given by the infinite sum of all negative energy levels
in~(\ref{eq:n=0energy})
\begin{equation*}
  E_0^{\mathrm{vac}} =
  - F - \frac{4\pi}{L}\sum_{l=1}^{\infty}
                        \sqrt{l^2+\left(\frac{FL}{2\pi}\right)^2}
  \;.
\end{equation*}
A simple method to deal with this sum is to change square roots to
powers of $d/2$ and use zeta function regularization (see,
eg.~\cite{elizalde89:_expres_casim_errat,elizalde89:_expres_casim})
one gets
\begin{equation}\label{eq:n=0vacE}
  E_0^{\mathrm{vac}} =
    \frac{F^2L}{8\pi^{3/2}}\Gamma\left(-\frac{d+1}{2}\right)
    +\sqrt{\frac{2F}{\pi L}}\e^{-FL}
  \;,
\end{equation}
where $d$ is 1.  The first term is just the normal infinite vacuum
energy density for massive field, and should be taken care of by
normal ordering of the operators in quantization, and the second one
is the Casimir force.

Analogous calculation in $n=1$ using energy
levels~(\ref{eq:n=1energy}) leads to the sum
\begin{equation*}
  E_1^{\mathrm{vac}} =
  -\frac{4\pi}{L}\sum_{l=1}^{\infty}
                    \sqrt{\left(l-\half\right)^2
                          +\left(\frac{FL}{2\pi}\right)^2}
  \;.
\end{equation*}
This again can be computed in a zeta function regularization style
(using eg.~\cite{Abbassi:1998hh})
\begin{equation}\label{eq:n=1vacE}
  E_1^{\mathrm{vac}} =
    \frac{F^2L}{8\pi^{3/2}}\Gamma\left(-\frac{d+1}{2}\right)
    -\sqrt{\frac{2F}{\pi L}}\e^{-FL}
  \;.
\end{equation}
Subtracting (\ref{eq:n=1vacE}) from (\ref{eq:n=0vacE}) we get for the
difference of vacuum energies in different gauge vacua
\begin{equation}\label{eq:deltae}
  \Delta E^{\mathrm{vac}} = E_1^\mathrm{vac}-E_0^\mathrm{vac} =
    -2\sqrt{\frac{2F}{\pi L}}\e^{-FL}  
  \;.
\end{equation}
We see, that the infinite contribution cancels exactly, and the finite
difference goes to zero exponentially with $L$.  Thus, we conclude
that in the limit of infinite space there is no energy difference
between different vacua, despite of na\"ively different fermionic
energy levels. As $\Delta E^{\mathrm{vac}}<0$ for finite system size,
the odd bosonic vacua are indeed the real vacua!

Note, that exactly the same result~(\ref{eq:deltae}) can be obtained
using Pauli-Villars regularization scheme also.

\section{Fermion number of the $n=1$ vacuum}
\label{sec:appFnumber}

%
\newcommand{\beq}{\begin{equation}}
\newcommand{\eeq}{\end{equation}}
\newcommand{\bea}{\begin{eqnarray}}
\newcommand{\eea}{\end{eqnarray}}

We calculate here the fermion number in the $n=1$ vacuum by different
means, starting from its definition.

The fermionic Lagrangian is invariant under the following global
transformations:
\begin{align*}
\Psi&\to e^{i\theta}\Psi,\\
\Psi^\dagger&\to e^{-i\theta}\Psi^\dagger.
\end{align*}
The conserved Noether current is
$j^\mu=\overline{\Psi}\gamma^\mu\Psi$, and the related charge is the
fermionic number $N_f=\int j^0 dx =\int\Psi^\dagger \Psi dx$. However,
if we quantize the system ($\Psi$ becomes operator and $N_f$ needs
normal ordering, $N_f=\frac{1}{2}\int\left(\Psi^\dagger
\Psi-\Psi\Psi^\dagger \right) dx$) the current is not conserved any
more, it suffer from the following anomaly:
\[
\partial_\mu j^\mu=\frac{e}{4\pi}\varepsilon^{\mu\nu}F_{\mu\nu}.
\]
The fermionic number vary in time as 
\[
\Delta N_f=\int \frac{e}{4\pi}\varepsilon^{\mu\nu}F_{\mu\nu} d^2x=\frac{e}{2\pi}\oint A\cdot dl.
\]
In the $A_0=0$ gauge, if we start with $N_f=0$ in vacuum $|0\rangle$,
then $N_f=0+\Delta N_f=\int A_1(x) dx=1/2$ in the sphaleron
configuration and $N_f=1$ in the vacuum $|1\rangle$. This result is
what we expect from the level-crossing picture.

These results may also be found by explicit calculations. The
sphaleron (kink) case was done eg.\ in the Chapter 9 of
\cite{Rajaraman}. In short: In the background of the sphaleron we have
one zero-mode for $\Psi$ and the other modes come in pairs (particle
and anti-particle):
\beq
\Psi(x,t)=b_0f_0(x)+\sum_{r=1}^\infty b_r e^{-iE_r t}
f_r^+(x)+\sum_{r=1}^\infty d_r e^{iE_r t} f_r^-(x).\label{psi}
\eeq
Imposing equal time anticommutating relations
\[
\{\Psi_\alpha
(x,t),\Psi^\dagger_\beta (y,t)\}=\delta_{\alpha\beta}\delta(x-y)
\]
and setting
other anticommutators to zero, we get for the operators $b,~d$:
\begin{align}
  \{b_r, b_{r'}^\dagger\}=\{d_r, d_{r'}^\dagger\}&=\delta_{rr'} \notag\\
  \{b_0,b_0^\dagger\}&=1 \label{com}
\end{align}
and all other anticomutators vanishes.
We can calculate the fermion number with (\ref{psi}) and (\ref{com}),
\bea
N_f&=&\frac{1}{2}\int\left(\Psi^\dagger \Psi-\Psi\Psi^\dagger \right) dx\notag\\
&=&b_0^\dagger b_0-\frac{1}{2} +\sum_{r=1}^\infty\left( b_r^\dagger b_r-d_r^\dagger d_r\right).\label{nf}
\eea
Application of the operator $N_f$ to the sphaleron configuration with
the zero-mode occupied gives $N_f(b_0^\dagger|0\rangle)=1/2$. Whereas
in the case of empty zero energy state: $N_f|0\rangle=-1/2$ (the
strange term $-\frac{1}{2}$ in (\ref{nf}) arise because we have a
single state. Such $\frac{1}{2}$-terms arise for each creation
operators, but they cancels between particle $b$ and antiparticle
$d$), In any vacua $|n\rangle$ each states of negative energy (created
by $d_r,~r=1,2,...$) correspond to a positive energy state (created by
$b_r^\dagger,~r=1,2,...$). The field is
\[
\Psi(x,t)=\sum_{r=1}^\infty b_r e^{-iE_r t} f_r^+(x)+ d_r e^{-iE_r t} f_r^-(x),
\]
where the $E_r$ and the $f_r$ depends on the topological number of the
vacuum. The fermion number is simply
$$N_f=\sum_{r=1}^\infty\left( b_r^\dagger b_r-d_r^\dagger d_r\right).$$
In particular $N_f|1\rangle=0$, $N_f b_1^\dagger|1\rangle=1$, as in usual vacua.

\section{Antiinstanton determinant}
\label{sec:detcalculation}

The determinant of the fermionic fluctuations around the
anti-instanton $\det'[K^\dagger K_{n=-1}]$ has been computed in
Ref. \cite{Burnier:2005he}. We need here the same determinant in the
background of the instanton ($n=1$). Noticing that $K^\dagger
K_{n=1}=KK^\dagger_{n=-1}$ allows for better comparison between these
two calculations. We may compare the operators $KK^\dagger_{n=-1}$ and
$K^\dagger K_{n=-1}$: they have the same spectrum
$\{\lambda_n\}_{n\neq 0}$ except that $K^\dagger K$ has a
supplementary mode with eigenvalue $\lambda_0=0$.  The determinant of
$\det[K^\dagger K_{n=-1}]$ normalized to vacuum looks like
$$\frac{\det[K^\dagger K_{n=-1}]}{\det[K^\dagger K_{vac}]}=\frac{\lambda_0\lambda_1...}{\lambda^{vac}_0\lambda^{vac}_1...}.$$
Removing the zero mode and inserting the value for the lowest
eigenvalue in the vacuum $\lambda^{vac}_0=F^2$ lead to:
$$\frac{\det'[K^\dagger K_{n=-1}]}{\det[K^\dagger K_{vac}]}=\frac{1}{F^2} \frac{\lambda_1\lambda_2...}{\lambda^{vac}_1\lambda^{vac}_2...}.$$
Naively we can guess that in the continuum limit, the eigenvalues in
the vacuum are close to each other and
\beq
 \frac{\det'[K^\dagger K_{n=-1}]}{\det[K^\dagger K_{vac}]}\sim\frac{1}{F^2}\frac{ \lambda_1\lambda_2...}{\lambda^{vac}_0\lambda^{vac}_1...}=\frac{1}{F^2}\frac{\det[K^\dagger K_{n=1}]}{\det[K^\dagger K_{vac}]}.\label{det}
\eeq
An explicit computation is performed in the following, and shows that
this naive expectation is correct in the cases of interests, even if
no general proof was found.

The computation of $\det[K^\dagger K_{n=1}]$ differ from the
calculation of $\det[K^\dagger K_{n=-1}]$ by the very fact that the
radial equations for the $\Psi^m_{L,R}$ are not
diagonal\footnotemark{}
in partial wave space (compare equation (46)
of ref. \cite{Burnier:2005he}):
\begin{widetext}
\begin{align}
  \left[\frac{\partial^2}{\partial r^2}
    +\frac{1}{r}\frac{\partial}{\partial r}
    -\frac{m^2}{r^2}-F^2f^2(r)+\frac{e}{2}(A'(r)+\frac{A(r)}{r})
    -\frac{e^2}{4}A^2(r) - me\frac{A(r)}{r}
  \right]  & \Psi_L^m
  \notag\\
  +\left[f\left(f'(r)-\frac{1}{r}f(r)
     -eA(r)f(r)\right)
   \right] & \Psi_R^{m-2} =0
  \;,\label{s1}\\
  \left[f\left(f'(r)-\frac{1}{r}f(r)-eA(r)f(r)\right)\right]
    & \Psi_L^{m}
  \notag\\
  +\left[\frac{\partial^2}{\partial r^2}
     +\frac{1}{r}\frac{\partial}{\partial r}-\frac{(m-2)^2}{r^2}
     -F^2f^2(r)+\frac{e}{2}(A'(r)+\frac{A(r)}{r})
     -\frac{e^2}{4}A^2(r)+\frac{(m-2)eA(r)}{r}
   \right]
    & \Psi_R^{m-2} = 0
  \;.\label{s2}
\end{align}
\end{widetext}
\leavevmode
\footnotetext{One is tempted to define a new numbering of
  the variables to put this matrix in a block diagonal form, however
  it means that we commute lines at infinity, which is not
  permitted. Moreover it is not clear how to rearrange the
  corresponding variables for the vacuum operator.}
Let us rename $\Psi_L^m=\psi_{2m}$ and $\Psi_R^m=\psi_{2m+1}$ and
define the operator $M_{ij}$ so that previous equations
(\ref{s1},\ref{s2}) are rewritten shortly as $M_{ij}\psi_j=0$. As in
equation (47, 48) of ref. \cite{Burnier:2005he}, the determinant can
be extracted from the solution of the following differential systems:
\[
M_{in} \psi_{nj}(r)=0,\quad M_{jj}^{vac}\psi^{vac}_j(r)=0,
\]
with boundary conditions
$$\lim_{r\to 0}\frac{\psi_{ij}(r)}{\psi^{vac}_i(r)}=\delta_{ij}.$$
The determinant is then given by 
\[
\det\left[\frac{\psi_{ij}(R)}{\psi^{vac}_i(R)}\right].
\]
The non zero elements of the matrice
$\frac{\psi_{ij}(R)}{\psi^{vac}_i(R)}=a_{ij}$ are on the diagonal or
of the form $a_{2i-3,2i}$, $a_{2i,2i-3}$, for any integer $i$.  Its
determinant can be computed with the following formula:
\[
\det[a_{ij}]=\prod_{i=-\infty}^\infty\left( a_{2i,2i}a_{2i-3,2i-3}-a_{2i,2i-3}a_{2i-3,2i}\right)\;.
\]
Note that there is no zero-mode in $K^\dagger K_{n=1}$ and its
regularization and renormalization is carried out like in
\cite{Burnier:2005he}.  The results of the numerical computation agree
to $10^{-3}$ accuracy to the formula (\ref{det}). An analytical
calculation is possible only in very simplified situations. We were
able to check formula (\ref{det}) for a modified instanton with
profile
$$A(r)=\frac{1}{r}\theta(r-a),\quad f(r)=\theta(r-a).$$
The computation is lengthy and will not be given here.

\bibliographystyle{h-physrev4}
\bibliography{onefermion,all}

\end{document}